\documentclass[10pt]{article} 
\usepackage[preprint]{main}

\usepackage{amsmath,amsfonts,bm}

\def\eqref#1{equation~\ref{#1}}

\def\1{\bm{1}}

\DeclareMathAlphabet{\mathsfit}{\encodingdefault}{\sfdefault}{m}{sl}
\SetMathAlphabet{\mathsfit}{bold}{\encodingdefault}{\sfdefault}{bx}{n}

\newcommand{\supp}[1]{\textcolor{black}{#1}}
\usepackage{url}
\usepackage{hyperref}
\usepackage{orcidlink}
\usepackage{cleveref}
\usepackage{multirow}
\usepackage{wrapfig}
\usepackage{listings}
\usepackage{xcolor}
\usepackage{booktabs}
\title{
\centering
Automatic Method Illustration Generation for AI Scientific Papers via Drawing Middleware Creation, Evolution, and Orchestration}

\author{
\centering
Zhuoling Li$^{1*}$, Jiarui Zhang$^{1*}$, Ping Hu$^{2}$, Jason Kuen$^{3}$, \\Jiuxiang Gu$^{3}$, Hossein Rahmani$^{1}$, Jun Liu$^{1}$\\
$^{1}$Lancaster University\\
$^{2}$University of Electronic Science and Technology of China\\
$^{3}$Adobe Research\\
{\small $^{*}$Equal contribution}
}

\def\month{MM}  
\def\year{YYYY}

\begin{document}

\maketitle

\begin{abstract}
Method illustrations (MIs) play a crucial role in conveying the core ideas of scientific papers, yet their generation remains a labor-intensive process. 
  Here, we take inspiration from human authors' drawing practices and correspondingly propose \textbf{FigAgent}, a novel multi-agent framework for high-quality automatic MI generation. Our FigAgent distills drawing experiences from similar components across MIs and encapsulates them into reusable drawing middlewares that can be orchestrated for MI generation, while evolving these middlewares to adapt to dynamically evolving drawing requirements. Besides, a novel Explore-and-Select drawing strategy is introduced to mimic the human-like trial-and-error manner for gradually constructing MIs with complex structures.
  Extensive experiments show the efficacy of our method.
\end{abstract}

\section{Introduction}
\label{sec:intro}

Method illustration (MI), a figure that conveys the workflow and core design insights of a proposed method, has become an indispensable component of modern AI scientific papers. Yet, creating high-quality MIs remains labor-intensive, often requiring hours or even days of iteration. 
Hence, automating this process is highly desirable and has attracted growing research attention~\cite{chang2025sridbench,liang2025diagrameval}.

To enable automatic MI generation, a few recent attempts~\cite{ zhu2026autofigure, zhu2026paperbanana, zaladiagrammergpt, rodriguez2023figgen} leverage image generation techniques~\cite{songdenoising, nichol2021improved, openai2023dalle3,openai2024gpt4o,google2025gemini3proimage} (e.g., Nano Banana Pro~\cite{google2025gemini3proimage}) to generate MIs from textual prompts. 
Yet, despite the demonstrated performance in natural or artistic image synthesis, existing image generation techniques are shown to still remain ill-suited for producing high-quality MIs~\cite{zuo2025nano, zhang-etal-2025-strict, chang2025sridbench,zhang2026sciflow,liang2025diagrameval}, due to two specific requirements of MIs: \emph{high-fidelity} and \emph{SVG-format}.

Firstly, for \textbf{high-fidelity}, unlike common natural or animated image generation, which typically involves generative models hallucinating or inferring additional visual details to enhance perceptual plausibility and aesthetics~\cite{chen2025survey, zuo2025nano}, MI generation aims to faithfully present the core ideas of a paper. This demands accurate visualization of concepts described in the paper (e.g., designed network modules), clear layout allocation, and precise rendering of symbolic and textual elements (e.g., concept names and equations), thereby imposing specific requirements on image fidelity that differ from those of conventional image generation tasks~\cite{zhang-etal-2025-strict,chang2025sridbench}. 
Achieving such fidelity can be particularly challenging due to the inherent characteristics of MIs.
Specifically, AI research papers typically involve multiple core concepts, each of which encodes complex semantics. To clearly convey these ideas, MIs often exhibit highly complex structures, usually comprising multiple visual modules that themselves contain rich internal substructures. Such intricate composition necessitates fine-grained and precise control over the generation process. 
Moreover, the research community evolves rapidly, and scientific papers frequently introduce new concepts. Such dynamics further increase the difficulty of accurately depicting core concepts in MIs and require the generation system to keep aligned with the rapidly evolving research community.
Besides fidelity, MIs are increasingly expected in \textbf{Scalable Vector Graphics (SVG) format}, which preserves visual quality under arbitrary scaling and supports flexible post-editing, and this practice has become increasingly common in the research community. Yet, image-generation-based methods~\cite{zhu2026paperbanana, zaladiagrammergpt,rodriguez2023figgen, openai2023dalle3, google2025gemini3proimage} usually produce resolution-limited and hard-to-edit raster images. 
This limitation is difficult to overcome through straightforward solutions such as converting raster outputs into SVGs via post-processing~\cite{carlier2020deepsvg, zhu2026autofigure,hu2024supersvg}, which still inherit the fidelity limitations of the underlying generation models and even often introduce additional artifacts, leading to fragmented and hard-to-edit primitives~\cite{zhang2024text,wang2025layered}.

To this end, automating the creation of high-fidelity SVG-format MIs still remains a very challenging and largely underexplored problem. Nevertheless, we observe that human authors are still able to produce high-quality MIs, albeit with substantial time investment. This naturally motivates us to reflect on how human authors draw MIs. 
\textbf{Firstly}, although MIs exhibit intricate compositions and can vary substantially in overall appearance across different papers, they can still share similar visual components at a finer granularity. 
For example, distinct model architectures may include common components (e.g., encoder–decoder structures and attention blocks). We find some authors will store these common components from previously created MIs as templates (which we call \emph{drawing middlewares} here) in their personal repositories. 
In this way, authors can construct complex new MIs through lightweight adjustments and assembly of these previously stored middlewares, thereby substantially simplifying and facilitating the challenging MI creation process.
\textbf{Besides}, we also observe that authors often create new component middlewares as new concepts emerge and also continuously evolve (update) existing ones, to keep them aligned with the dynamically evolving research community.
\textbf{Moreover}, when handling complex MIs, instead of constructing the entire figure in one go, authors often draw in a trial-and-error manner, by tentatively adding a few strokes to see how the figure takes shape. They usually selectively proceed when the current attempt appears promising; otherwise, they revert to a previously satisfactory version and explore alternative drawing directions.
\textbf{Lastly}, authors typically rely on professional drawing software to construct MIs, which usually supports exporting figures in SVG format.

Through the above drawing practices, human authors can well handle the inherent characteristics of MIs and effectively produce high-fidelity SVG-format MIs.
Inspired by this, we propose \textbf{FigAgent}, the first multi-agent framework that automatically mirrors humans' practices in creating, evolving, and orchestrating (assembling) reusable drawing middlewares for drawing high-fidelity SVG MIs of AI scientific papers.
Specifically, our framework drives a vision-LLM-powered agent to automatically collect MIs from online published papers, including those from top-tier venues (e.g., ECCV and CVPR), to construct an \emph{experience dataset}, from which it distills common visual components and encapsulates them into a \textbf{drawing middleware repository}. 
Notably, each middleware is implemented as \emph{an executable Python function} corresponding to a DrawIO-supported drawing operation, parameterized by attributes such as position and size. For example, a middleware \texttt{Attention\_map(x, y, w, h, "diagonal")} places a diagonal attention map at location $(x, y)$ with width $w$ and height $h$ on the canvas. 
During MI generation, our framework employs another agent that interacts with professional drawing software (e.g., DrawIO) and autonomously orchestrates appropriate middlewares to draw complex MIs. 
Notably, since these middlewares are executable Python functions that encapsulate SVG code, MIs generated by the agent can be seamlessly rendered into SVG format and remain readily editable. 
Besides, inspired by Darwin's theory of evolution and Genetic Programming~\cite{fisher1999genetical,koza1994genetic, ma2025automated, zhang2025evoflow}, our framework continuously evolves the middleware repository to improve its overall efficacy and keep it aligned with newly emerging concepts in the rapidly evolving research community.

More importantly, given the considerable complexity of MIs, which usually comprise multiple interrelated visual modules whose individual depiction and joint spatial arrangement critically affect the overall figure quality, our FigAgent further introduces \textbf{Explore-and-Select}, a novel drawing strategy that mirrors human authors' trial-and-error drawing behavior to facilitate effective middleware orchestration on the canvas. Our Explore-and-Select strategy models the drawing process as a search tree~\cite{hart1968formal, browne2012survey, kocsis2006bandit, marsland1986review} and guides the agent to explore proper middleware orchestration manners, encouraging the generation process to progress along promising paths.

Our contributions are summarized as follows. 
1) Our FigAgent mimics the middleware creation, evolution, and orchestration practices of human authors, enabling the direct generation of high-fidelity SVG MIs for AI scientific papers.
2) By creating and evolving drawing middlewares through agent-driven collaboration, and by orchestrating them under the novel Explore-and-Select drawing strategy, our framework can effectively handle complex MIs while remaining aligned with the rapidly evolving research community.
3) Our method achieves state-of-the-art performance on evaluated benchmarks.

\section{Related Work}
Recent advances in generative models for natural language~\cite{achiam2023gpt,minaee2024large} and natural images~\cite{google2025gemini3proimage,nichol2021improved} have inspired growing interest in automatic scientific illustration generation. Existing approaches~\cite{zaladiagrammergpt,rodriguez2023ocr,chang2025sridbench,belouadi2024detikzify,belouadi2025tikzero,belouadiautomatikz,wei2025words,yangomnisvg} can be categorized into two paradigms: (1) methods that rely on image generation models, and (2) methods that generate code that can be seamlessly converted into SVGs.

Some works~\cite{zaladiagrammergpt,rodriguez2023ocr,chang2025sridbench,rodriguez2023figgen} utilize image-generation models (e.g., diffusion models) to produce illustrations for scientific papers. For instance, FigGen~\cite{rodriguez2023figgen} and OCR-VQGAN~\cite{rodriguez2023ocr} train text-to-image models to generate pixel-based illustrations from textual prompts, while 
AutoFigure~\cite{zhu2026autofigure} and PaperBanana~\cite{zhu2026paperbanana} further employ LLMs to parse paper texts before invoking image generation models.
To enhance editability, some works~\cite{EditBanana2025,hu2024supersvg} introduce post-processing techniques. EditBanana~\cite{EditBanana2025} uses OCR and segmentation to detect and convert textual and basic graphic elements into editable formats, while SuperSVG~\cite{hu2024supersvg} adopts differentiable vectorization to convert raster images.
Yet, such post-hoc approaches still inherit the fidelity limitations of the underlying image-generation models, and the vectorized outputs often exhibit imprecise geometric relationships and irregular paths~\cite{belouadiautomatikz,wu2023iconshop}, limiting their suitability for high-quality MI generation.

Meanwhile, directly generating editable SVG illustrations usually better aligns with the needs of the research community. 
Along this line, some works~\cite{belouadi2024detikzify,belouadi2025tikzero,belouadiautomatikz,wei2025words} leverage LLMs to generate \LaTeX{} TikZ code for SVG-format illustrations, while others~\cite{rodriguez2025starvector,xing2025empowering,wu2023iconshop, wu2025chat2svg, wang2025svgen,yangomnisvg} explore generating XML-based code that can be directly rendered into SVGs. 
Despite the progress, these methods usually remain limited to producing either simple icons~\cite{wu2023iconshop,rodriguez2025starvector,xing2025reason,wu2025chat2svg,yangomnisvg,wang2025svgen} or basic scientific charts~\cite{belouadi2025tikzero,belouadi2024detikzify,wei2025words} (e.g., bar plots), and thus still struggle with MI generation.
A more recent attempt~\cite{guo2025paper2sysarch} explores generating code for MI layout, which, however, still relies on image generation models for rendering main visual modules, and thus the resulting illustrations remain limited to editing.

Differently, our FigAgent autonomously creates and evolves drawing middlewares to support rendering complex MIs, and introduces a novel Explore-and-Select strategy to guide the drawing process, which, to the best of our knowledge, for the first time enables the direct generation of high-fidelity SVG MIs.

\section{Proposed Method}
\label{sec:overview}
\begin{figure}[t]
  \centering
  \includegraphics[width=0.98\linewidth]{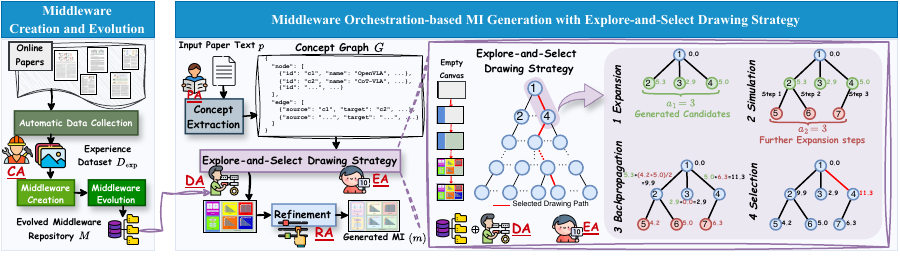}
  \vspace{-0.5cm}
  \caption{
Overview of our {FigAgent}. Our framework comprises five agents: CA, PA, DA, EA, and RA.
In the \emph{Middleware Creation and Evolution} stage, the CA
creates a set of middlewares $\mathcal{M}$ from the automatically collected $\mathcal{D}_{\text{exp}}$, and evolves $\mathcal{M}$ to improve its efficacy and maintain alignment with the evolving research community.
During \emph{Middleware Orchestration-based MI Generation}, the PA parses paper text $p$ into a concept graph $\mathcal{G}$. The DA then orchestrates middlewares from $\mathcal{M}$ to gradually render concepts onto the canvas, under the Explore-and-Select strategy (we show a toy case where $(a_1,a_2,\beta)$ is simply set to $(3,3,1)$ for ease of understanding), with the EA providing evaluation feedback. Lastly, the RA refines the result to produce the MI $m$.
}
  \label{fig:main}
  \vspace{-0.5cm}
\end{figure}

Given a paper text $p$ (typically the text of the introduction and method sections), our goal is to automatically generate a high-quality MI ($m$) that intuitively conveys the core ideas of $p$.
The MI $m$ is stored as XML-based DrawIO code, which can be exported to SVG format for visualization and can be readily edited by authors. As shown in~\cref{fig:main}, our framework comprises two key stages, both executed through the collaboration of LLM agents:
\textbf{1) Middleware Creation and Evolution.}
Our framework automatically and continuously collects publicly available AI scientific papers and constructs an \emph{experience dataset} $\mathcal{D}_{\text{exp}}=\{(p_n, m_n)\}_{n=1}^{N_{\text{exp}}}$, where $N_{\text{exp}}$ denotes the number of paper-MI pairs. Through multi-agent collaboration, reusable common visual components are identified and extracted from MIs in $\mathcal{D}_{\text{exp}}$, and then encapsulated into callable Python functions to construct a drawing middleware repository $\mathcal{M}$ via our dedicated middleware creation and evolution mechanisms, which can support subsequent MI generation.
\textbf{2) Middleware Orchestration-based MI Generation with Explore-and-Select.}
Equipped with $\mathcal{M}$, our framework parses the paper text $p$ to extract key concepts and orchestrates appropriate middlewares to render these concepts. This process is guided by a novel \emph{Explore-and-Select
} drawing strategy, under which our agents collaboratively perform drawing, evaluation, and refinement to generate high-quality MIs.

Below, we first present the agents in our framework in~\cref{sec:agents}, and then describe how $\mathcal{M}$ is created and evolved in~\cref{sec:middleware}.
Finally, we discuss the overall MI generation process guided by the Explore-and-Select strategy in~\cref{sec:mcts}.

\subsection{Agents in FigAgent}
\label{sec:agents}

We begin by analyzing the human drawing practices to inform the design of our framework. 
Before drawing, authors often accumulate drawing experiences by studying high-quality MIs from existing papers, and some authors even distill common visual components into reusable drawing templates (middlewares) to subsequently support the drawing of new MIs. When drawing for a given paper text, they usually first parse the text to identify key concepts and their relationships, then progressively render each concept onto the canvas through iterative trial-and-error revisions, and finally perform a refinement step to further adjust the overall MI, e.g., adding connections and tuning alignment. Motivated by this analysis, as shown in~\cref{fig:main}, our FigAgent framework mainly consists of the following five components (agents) that mirror humans' drawing workflow, \emph{with more details (including LLM prompts) provided in \supp{Supplementary}}.

\textbf{(1) Constructor Agent (CA):}
The vision-LLM-powered CA automatically scrapes publicly available paper texts and corresponding MIs, and filters out low-quality paper-MI pairs to construct the experience dataset $\mathcal{D}_{\text{exp}}$. Based on $\mathcal{D}_{\text{exp}}$, the CA then constructs and maintains a drawing middleware repository $\mathcal{M}$ to support MI generation. 
\noindent\textbf{(2) Parser Agent (PA):}
Given a paper text $p$, the LLM-powered PA first identifies the paper's research theme $t$, then extracts core concepts and organizes them into a concept graph $\mathcal{G}=(\mathcal{C},\mathcal{E})$, where $\mathcal{C}=\{c_i\}$ is the set of concept nodes and $\mathcal{E}$ encodes inter-concept relationships.
\noindent\textbf{(3) Drawer Agent (DA):}
The vision-LLM-powered DA interfaces with DrawIO and draws the concepts in $\mathcal{G}$ onto the canvas one by one. Notably, during this process, the DA will select appropriate middlewares from $\mathcal{M}$ and orchestrate them for drawing.
\noindent\textbf{(4) Evaluator Agent (EA):}
Powered by a vision-LLM, the EA assesses MI quality, producing textual revision feedback and a numeric quality score that serves as the reward signal in our Explore-and-Select strategy.
\noindent\textbf{(5) Refiner Agent (RA):}
Powered by a vision-LLM, the RA applies final adjustments to MIs, including adding inter-concept connections based on $\mathcal{G}$, tuning spacing, and harmonizing the overall visual style.

These agents collaboratively complete the two key stages of our framework for high-quality MI generation.  Specifically, the  CA, PA, DA, and EA participate in the creation and evolution process of the \emph{Drawing Middleware Repository}, while, guided by our Explore-and-Select drawing strategy, the PA, DA, EA, and RA collectively accomplish \emph{Middleware Orchestration-based MI Generation}.

\subsection{Middleware Creation and Evolution}
\label{sec:middleware}
Given the inherent complexity of MIs, faithfully drawing them from paper texts while ensuring convenient editability remains challenging for the DA. Nevertheless, we observe that 
although different MIs usually vary substantially in overall appearance, they can still share similar visual components at a finer granularity, presenting a key opportunity to mitigate this challenge.

Based on this, we propose a novel \emph{middleware-driven} MI generation paradigm. Our framework drives the CA to identify recurring visual components across MIs and encapsulate them into a drawing middleware repository $\mathcal{M}$, so that complex MI generation can be decomposed into the \emph{selection}, \emph{parameterization}, and \emph{orchestration} of well-established middlewares, greatly reducing drawing difficulty. Besides, leveraging the dual nature of SVG-format MIs-which can be perceived as both visual graphics and XML-based code-each middleware is implemented as an executable Python function that generates corresponding code for a reusable visual component. This design simultaneously ensures visual fidelity and convenient editability, and is naturally compatible with vision-LLM-powered agents that exhibit strong capabilities in both visual understanding and code generation~\cite{hurst2024gpt,yin2024survey}.
Moreover, our framework also drives the CA to continuously evolve $\mathcal{M}$ through LLM-simulated biological genetics, maintaining alignment with newly emerging concepts (e.g., novel network modules) in the rapidly evolving research community while enhancing overall effectiveness.

Below, we first describe how our framework initially creates $\mathcal{M}$, and then elaborate on the middleware evolution mechanism that makes $\mathcal{M}$ ready for use.

\noindent\textbf{Middleware Creation.}
To create $\mathcal{M}$, the CA first automatically collects publicly available paper-MI pairs to construct the experience dataset $\mathcal{D}_{\text{exp}}$.
By analyzing $\mathcal{D}_{\text{exp}}$, we obtain two key observations that inform the construction of $\mathcal{M}$:
\emph{(1)} MI presentation is closely related to the paper’s research theme (topic), and even the same concept can be visualized differently across themes. For example, a “feature map” is often depicted as a diagonal grid pattern in LLM reasoning papers, whereas in object detection papers, it may appear as a multi-scale feature pyramid.
\emph{(2)} Within the same research theme, similar concepts often tend to exhibit common visual components.
Given this, our framework organizes $\mathcal{M}$ in a three-level structure: \emph{theme} (e.g., object detection, image generation), \emph{concept} (e.g., Transformer), and \emph{component} (e.g., attention layers), to facilitate the DA in retrieving appropriate middlewares according to drawing requirements (detailed in~\cref{sec:mcts}).
Based on this structure, our framework can create middlewares at the component level for each theme–concept pair.

For each paper-MI pair $(p, m)$ in $\mathcal{D}_{\text{exp}}$, the PA first parses $p$ to obtain a set of core concepts $\mathcal{C}$ and identifies the paper theme $t$.
Then, for each concept $c$ in $\mathcal{C}$, the CA jointly analyzes both the visual SVG and XML-based DrawIO code of the MI $m$, with a focus on the regions relevant to $c$, and identifies a set of key components that constitute concept $c$ (detailed prompt is in \supp{Supplementary}).
More crucially, instead of storing these identified components as static code segments, the CA abstracts each into a parameterized middleware: it encapsulates the corresponding code segment as a callable Python function, while exposing key drawing attributes (e.g., position, size, and visual patterns) as adjustable function parameters with sensible defaults inferred from the original MI.
For example, a feature pyramid component is abstracted into a middleware \texttt{Feature\_Pyramid(x, y, w, h, num\_levels, shape\_mode)}, where \texttt{x, y, w, h} control its placement and size on the canvas, \texttt{num\_levels} specifies the number of pyramid layers, and \texttt{shape\_mode} (e.g., ``cuboid'', ``rectangle'') specifies the geometric form used to depict each pyramid level. 
In this way, a single middleware can flexibly adapt to diverse drawing contexts through proper parameter configuration. The resulting function-based middlewares are then indexed under the theme-concept pair $(t, c)$ and added into the initial middleware repository $\mathcal{M}$.

By repeating the above middleware creation process for all samples in $\mathcal{D}_{\text{exp}}$, our framework creates the initial middleware repository $\mathcal{M}$ defined as
\begin{equation}
\vspace{-3mm}
\mathcal{M}
=
\bigcup_{(t_i,c_j)\in T\times C}
\{(t_i,c_j,F_{ij})\},
\end{equation}
where $T=\{t_i\}_{i=1}^{N_T}$ and $C=\{c_j\}_{j=1}^{N_C}$ are sets of \emph{themes} and \emph{concepts}, and ${F}_{ij} = \{f_1, f_2, \ldots, f_{N_{ij}}\}$ denotes the set of middlewares associated with theme $t_i$ and concept $c_j$.
Yet, since middlewares are independently extracted from different papers, the initially created $\mathcal{M}$ may contain semantically overlapping concepts (e.g., ``self-attention module'' and ``multi-head attention'') under the same research theme, introducing redundancy.
To obtain a more compact repository, the CA further performs a \emph{merging} step: it computes semantic similarity among concepts within each theme, consolidates highly similar concepts into unified entries, and leverages the LLM to identify and remove functionally redundant middlewares among the merged concepts (more details in~\supp{Supplementary}).
The resulting consolidated $\mathcal{M}$ is then passed to the middleware evolution stage.

\noindent\textbf{Middleware Evolution.}
Due to the heterogeneous quality of MIs in $\mathcal{D}_{\text{exp}}$, the initially created $\mathcal{M}$ may still contain some low-quality middlewares, necessitating deletion or refinement.
Moreover, exploring and incorporating new middlewares into $\mathcal{M}$ is also essential to adapt to newly emerging concepts (e.g., novel model architectures) as the research community continues to evolve.
To this end, our framework drives the CA to periodically incorporate paper-MI samples from newly published papers into $\mathcal{D}_{\text{exp}}$, and, from these newly added samples, re-invoke the middleware creation mechanism aforementioned to create new middlewares and integrate them into $\mathcal{M}$.
Moreover, our framework also seeks to evolve the middlewares in $\mathcal{M}$ to enhance their overall efficacy. Yet, this is non-trivial, since unlike vectorized parameters in deep learning models, middlewares are implemented as Python functions and thus are not directly differentiable.

Drawing inspiration from evolutionary biology and genetic programming~\cite{koza1994genetic}, our framework employs a novel LLM-driven middleware evolution mechanism to optimize $\mathcal{M}$ in a non-gradient manner. 
Specifically, our framework conceptualizes each middleware $f_k$ in $\mathcal{M}$ as an individual, the initially created repository $\mathcal{M}$ as the population, and the experience dataset $\mathcal{D}_{\text{exp}}$ as the environment. The LLM-powered CA is then prompted to simulate a biological evolutionary process over this population, aiming to improve its \emph{fitness} to the environment, namely, the overall middleware efficacy. This process proceeds iteratively through two alternating steps: \emph{Middleware Evaluation}, which assesses the population's fitness, and \emph{Middleware Update}, which emulates biological evolutionary operators, including selection, crossover, and mutation, to optimize the repository, with the former guiding the latter and controlling the iterative process.

\underline{Middleware Evaluation.}
The middleware evaluation step measures the fitness of the population $\mathcal{M}$ within the environment $\mathcal{D}_{\text{exp}}$, serving as the objective function that informs the consequence of each middleware update step: guiding our framework to retain beneficial updates and discard detrimental ones.
A straightforward approach for measuring fitness is to evaluate the MI generation performance of our framework using $\mathcal{M}$. However, solely considering this can be one-sided, as $\mathcal{M}$ may contain low-quality middlewares that are rarely invoked during generation and thus have a negligible impact on the final MI quality, yet their presence can still mislead subsequent middleware update operations (e.g., being selected as parents for crossover). Our framework thus additionally evaluates each individual middleware's efficacy, guiding the update process to optimize the repository $\mathcal{M}$ at a finer granularity.
During MI generation, the DA selects and orchestrates middlewares to render concepts one by one (detailed in~\cref{sec:mcts}). This process naturally provides a signal for evaluating individual middleware efficacy: each time a middleware $f_k$ is used, the EA assesses the visualization quality of the rendered concept it contributes to (detailed in~\supp{Supplementary}). Our framework tracks the utilization of each middleware $f_k$ and defines the \emph{Middleware Efficacy Score} (MES) as $\mathrm{MES}_{f_k} = S_{f_k} / N_{f_k}$, where $S_{f_k}$ is the sum of quality scores across all concept renderings that $f_k$ contributes to and $N_{f_k}$ is its total number of utilizations. A higher MES reflects greater efficacy.
Based on this, we define the objective function that jointly incorporates generation-level and individual-level metrics:
\vspace{-0.3cm}
\begin{equation}
\label{eq:obj}
\mathcal{L} = \frac{1}{|\mathcal{P}|}\sum_{p_{\text{exp}} \in \mathcal{P}} q(p_{\text{exp}})
\;+\; 
\frac{1}{|\mathcal{M}_{\mathcal{P}}|}\sum_{f_k \in \mathcal{M}_{\mathcal{P}}} \mathrm{MES}_{f_k},
\vspace{-3.mm}
\end{equation}
where $\mathcal{P}$ denotes a batch of paper texts sampled from $\mathcal{D}_{\text{exp}}$, $q(p_{\text{exp}})$ is the quality score assessed by the EA for the MI generated from $p_{\text{exp}}$, and $\mathcal{M}_{\mathcal{P}} \subseteq \mathcal{M}$ is the subset of middlewares invoked during MI generation for $\mathcal{P}$. 
By incorporating both generation-level and individual-level evaluation, this objective function provides comprehensive guidance for the subsequent middleware update step.

\underline{Middleware Update.} 
In this step, our framework employs the CA to optimize $\mathcal{M}$ through three key operations analogous to biological evolution~\cite{fisher1999genetical,koza1994genetic}: selection, mutation, and crossover. Each operation is realized via carefully crafted prompts within the CA (provided in~\supp{Supplementary}), enabling it to refine or delete existing middlewares, and explore new ones.
\emph{(1) Selection.} Mimicking natural selection, where only the fittest individuals survive, the CA retains effective middlewares and removes ineffective ones based on their MES, thereby eliminating noise and preventing low-quality middlewares from propagating into subsequent operations.
\emph{(2) Mutation.} Analogous to genetic mutation that introduces phenotypic diversity within a species, the CA modifies a middleware function by adjusting its parameters or expanding its functionality, thereby inducing variations that encourage the exploration of potentially superior middleware designs.
\emph{(3) Crossover.} Analogous to genetic recombination that produces offspring inheriting advantageous traits from both parents, the CA generates new middlewares by combining complementary features from two or more selected middlewares, creating potentially more versatile and effective variants.

Overall, the middleware evolution process is guided by the objective function $\mathcal{L}$ in~\cref{eq:obj}, which is maximized to improve the overall efficacy of the repository. Middleware updates that reduce $\mathcal{L}$ are discarded via a rollback mechanism, and early-stopping is applied when no significant improvement is observed over consecutive iterations. The evolution process terminates once the stopping condition is met or the maximum number of iterations is reached.

\subsection{Middleware Orchestration-based MI Generation with Explore-and-Select}
\label{sec:mcts}

After creating and evolving the middleware repository $\mathcal{M}$ and equipping it for the DA, our framework should be ready for MI generation. A natural pipeline thus is to parse the paper text to extract key concepts via the PA, render them via the DA, and finally employ the RA to refine the resulting MI.
Besides, to further simplify complex MI generation and improve the quality of generated MIs, our framework mirrors the common practice of human authors, who usually draw MIs progressively rather than constructing the entire figure in one go. 
Based on this intuition, our framework drives the DA to render one concept at a time on the canvas, gradually completing the entire MI. 
We detail how the DA orchestrates middlewares from $\mathcal{M}$ to render each concept later.

However, even with $\mathcal{M}$ and progressive drawing, we observe that this pipeline can still yield sub-optimal MIs. This is mainly because the LLM-powered DA is inherently stochastic: when rendering the same concept, it can produce substantially different visual outcomes by choosing different middlewares or even applying the same middlewares with different parameterization. We refer to these variations as distinct \emph{drawing choices}. 
More importantly, since an MI comprises multiple interrelated concepts whose individual depiction and joint spatial arrangement critically determine the overall figure quality, drawing choices are path-dependent: the choice made for one concept constrains how subsequent concepts can be depicted and arranged. Thus, inappropriate early choices can propagate through the rendering sequence and gradually degrade the final MI.

To tackle this, our framework introduces a novel Monte Carlo Tree Search~\cite{browne2012survey,marsland1986review} inspired drawing strategy, namely \textbf{Explore-and-Select},
which first \emph{explores} multiple candidate drawing choices and evaluates their long-term effects on the MI, and then \emph{selects} the most promising one to advance the MI, thereby alleviating premature commitment to suboptimal choices and enabling the discovery of more coherent and synergistic combinations of drawing choices.

Below, we first detail how the DA selects and orchestrates middlewares to render each concept, and then elaborate on how the Explore-and-Select strategy guides concept rendering towards promising paths.

\noindent\textbf{Concept Rendering.} 
During MI generation, the DA selects and orchestrates appropriate middlewares from $\mathcal{M}$ to sequentially render the concepts $\mathcal{C}=\{c_i\}$ extracted from the paper text $p$. This process can be formulated as:
\begin{equation}
\label{eq:invok}
    m^{(i)} = \texttt{Drawer}(t, c_i, m^{i-1}; \mathcal{M}, H),
\end{equation}
where $m^{(i)}$ denotes the generated code segment for rendering concept $c_i$, $m^{i-1}$ is the accumulated code after the first $i{-}1$ concepts have been rendered, $t$ is the research theme inferred from $p$, and $H$ denotes the historical middleware usage experiences.
Concretely, this rendering process involves two steps: \emph{retrieval} and \emph{filtering}.
In \emph{retrieval}, our framework computes the semantic embedding similarity between the target concept $c_i$ and concepts under the same theme $t$ in $\mathcal{M}$, and collects middlewares associated with the top-$K$ most similar concepts to form a candidate set $F_{c_i}$.
Since such retrieval may introduce irrelevant or redundant middlewares due to coarse semantic embedding matching, a subsequent \emph{filtering} step prompts an LLM to prune $F_{c_i}$, and retain only the most suitable middlewares (detailed prompt in~\supp{Supplementary}).
The resulting filtered set, denoted as $F'_{c_i}$, together with $H$, is then provided to the DA for concept rendering. More details about the concept rendering process are in~\supp{Supplementary}.

\noindent\textbf{Explore-and-Select Strategy.}
To guide the DA to select promising drawing choices, as shown in~\cref{fig:main}, our Explore-and-Select strategy formulates the MI generation process as a search tree~\cite{ browne2012survey, kocsis2006bandit}. Each node represents an intermediate canvas state (i.e., a partially constructed MI), with the root node denoting an empty canvas. Each edge denotes a drawing choice made by the DA that advances the MI by drawing the next concept through the aforementioned concept rendering process.
This strategy iteratively builds a promising path through the tree. At each iteration, it first \emph{expands} the current node by exploring multiple candidate drawing choices, producing a set of child nodes, then \emph{simulates} further expansions from these child nodes to score the long-term impact of these drawing choices, \emph{backpropagates} the resulting scores to update their evaluations, and finally \emph{selects} the most promising child node to advance the MI. This four-step cycle repeats until all concepts are completed. We detail each step below.

\underline{Expansion.}
Starting from the terminal node of the current path (node 1 in~\cref{fig:main}), or the root node (empty canvas) in the first iteration, the DA renders the next concept using $a_1$ different drawing choices, producing $a_1$ candidate MI states (child nodes; i.e., nodes 2-4 in~\cref{fig:main}). These child nodes are then scored by the EA and ranked.  
Since each child node extends its parent node by completing one additional concept, we define the \emph{node score} as the incremental gain in MI quality score relative to its parent, rather than the absolute score assessed by the EA. Accordingly, the root node (empty canvas) is assigned a node score of zero. Notably, if all child nodes receive node scores below a predefined threshold, the DA regenerates $a_1$ children by incorporating the EA's revision feedback.

\underline{Simulation.}
While the above node scores can reflect the immediate impact of drawing choices, they still fail to capture their effects on future compositions.
To this end, we simulate further $a_2$  expansions from these $a_1$ generated nodes to estimate the long-term impacts of drawing choices.
At each expansion step, to prevent the agent from being trapped in local optima by only selecting the currently best-performing nodes, we also retain exploration opportunities for choices that exhibit lower initial scores but may still possess latent potential. Specifically, our framework introduces a UCT-inspired criterion~\cite{kocsis2006bandit} that balances exploitation (favoring nodes with higher estimated values) and exploration (encouraging less-explored nodes) by computing the \emph{UCT value} for the nodes:
$\mathrm{UCT}_j =\frac{2.0}{\mathrm{rank}_j+1}+\beta \sqrt{\frac{2.0}{\mathrm{child\_count}_j+1}},$
where $j$ indexes a node, $\mathrm{rank}_j$ denotes its relative rank based on the initial node scores, $\mathrm{child\_count}_j$ is the number of child nodes already expanded from node $j$, and $\beta$ controls the exploration--exploitation trade-off. 
At each step, the DA advances the node with the highest UCT value by rendering one subsequent concept, generating a new child node.

\underline{Backpropagation.}
To incorporate the long-term estimates obtained during simulation into the current decision, the scores of newly generated child nodes (nodes 5, 6, 7 in~\cref{fig:main}) are backpropagated through the tree. Concretely, a parent node's score is updated by adding the average score of its child nodes. 
In this way, each drawing choice is assessed not only by its immediate impact on the MI but also by its long-term effects on possible future drawing paths.

\underline{Selection.}
The node with the highest updated score (node 4 in~\cref{fig:main}) is selected as the new terminal node for the next iteration. 

The above four-step cycle will repeat until all concepts are completed. We provide more details about the Explore-and-Select strategy in~\supp{Supplementary}.

\subsection{Preparation and Testing.}
\noindent\textbf{Preparation.} 
Prior to MI generation, the CA automatically
constructs the middleware repository $\mathcal{M}$ from the collected experience dataset $\mathcal{D}_{\text{exp}}$ as detailed in~\cref{sec:middleware}. By automatically and periodically incorporating newly published papers and invoking the middleware creation and evolution mechanisms, the CA keeps $\mathcal{M}$ ready for use and aligned with the evolving research community.
\quad
\noindent\textbf{Testing.} 
Given a paper text $p$, the PA first parses it into a concept graph $\mathcal{G}$. The DA then renders concepts in $\mathcal{G}$ by orchestrating middlewares from $\mathcal{M}$ under the Explore-and-Select strategy, with the EA providing evaluative guidance during this process. Finally, the RA performs refinement to obtain the final MI $m$.

\section{Experiments}

\begin{figure*}[t]
  \centering
  
 \includegraphics[width=1\linewidth]{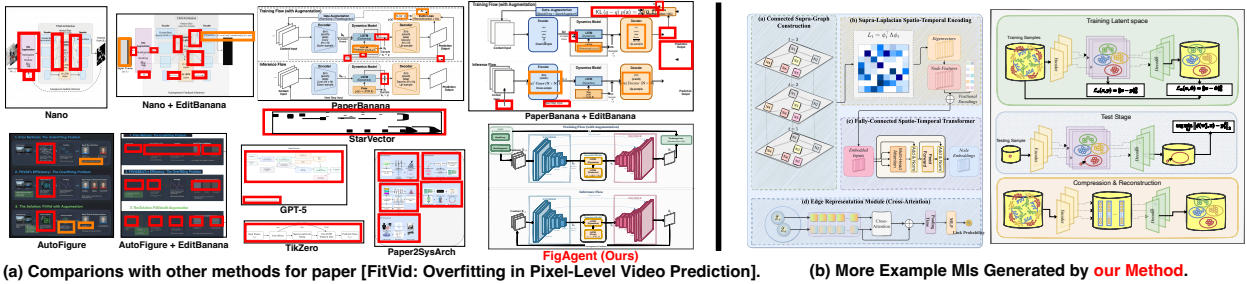}
  \vspace{-0.5cm}
  \caption{
  Qualitative Results. \textcolor{red}{Red boxes} denote structural defects, e.g., layout inconsistencies and missing components. \textcolor{orange}{Orange boxes} denote low-fidelity details like blurriness. Zoom in for better view. \textbf{More  zoomed-in examples are in \supp{Supplementary}}.
   }
  \label{fig:vis}
  \vspace{-0.5cm}
\end{figure*}

\noindent\textbf{{Implementation Details.}} 
We use \texttt{GPT-5}~\cite{openai2026gpt5} to power our agents, with evaluations using other LLM backbones in~\supp{Supplementary}. We use \texttt{all-mpnet-base\\-v2}~\cite{all-mpnet-base-v2} to encode concepts into embeddings for middleware retrieval.
More implementation details (including agent prompts) are in~\supp{Supplementary}.

\noindent\textbf{Dataset.}
Following the data collection pipeline of~\cite{liang2025diagrameval}, our framework collects MIs and corresponding paper texts (from major venues such as ECCV and CVPR) published up to December 2025, and extracts the XML code from the MI files, yielding 4,964 paper–MI (XML code) pairs in total. We randomly sample 80\% of the pairs to build $\mathcal{D}_\text{exp}$, while the remaining 20\% serve as the test set (named \textbf{FigAgentBench}) for evaluation.
We also conduct evaluation on PaperBananaBench~\cite{zhu2026paperbanana}.
More dataset details are in~\supp{Supplementary}.

\noindent\textbf{Evaluation Metrics.}
We use DiagramEval~\cite{liang2025diagrameval}, a metric designed for assessing AI-generated MIs, which conceptualizes the generated MI and its corresponding paper text as graphs, and evaluates their node alignment (\textbf{NA}) and path alignment (\textbf{PA}).
Besides, we use the Vision-LLM evaluator in~\cite{zhu2026paperbanana} as a complementary metric that assesses MI quality and reports an overall quality score, denoted as \textbf{VLM}. 
More details about evaluation metrics are  in~\supp{Supplementary}.

\subsection{Evaluation on MI Generation}

\noindent\textbf{Baseline Methods.}
We compare our framework against two categories of baselines.
\emph{(1) Image-based Methods}, which employ image generation models as the core component for MI generation. We consider Nano Banana Pro (Nano)~\cite{google2025gemini3proimage}, which is prompted to produce raster MIs in an end-to-end manner. This category also includes AutoFigure~\cite{zhu2026autofigure} and PaperBanana~\cite{zhu2026paperbanana}, which additionally leverage LLMs to parse paper texts before invoking image generation models.
\emph{(2) SVG-based Methods}, which can directly generate easily editable SVG illustrations. We consider GPT-5~\cite{openai2026gpt5}, which is prompted to generate XML-based DrawIO code from paper texts, as a representative baseline. We also include TikZero~\cite{belouadi2025tikzero} and StarVector~\cite{rodriguez2025starvector}, which fine-tune LLMs to generate TikZ or XML code for producing SVGs,
as well as Paper2SysArch~\cite{guo2025paper2sysarch}, which employs agents to generate XML code.
Besides, we also utilize EditBanana~\cite{EditBanana2025}, a widely adopted raster-to-SVG tool, to transform the outputs of image-based methods into editable SVGs for further comparison. 
More details about these baselines are in~\supp{Supplementary}.

\setlength{\columnsep}{0.05in}
\begin{wraptable}[10]{r}{0.6\columnwidth}
\centering
\vspace{-0.8cm}
\caption{Quantitative comparisons of different methods on MI generation quality.
}
\resizebox{1\linewidth}{!}{
\begin{tabular}{l ccc ccc c| ccc ccc c}
\toprule
\multirow{3}{*}{\textbf{Method}}
& \multicolumn{7}{c|}{\textbf{FigAgentBench}}
& \multicolumn{7}{c}{\textbf{PaperBananaBench~\cite{zhu2026paperbanana}}} \\
\cmidrule(lr){2-8} \cmidrule(lr){9-15}
& \multicolumn{3}{c}{\textbf{NA}$\uparrow$}
& \multicolumn{3}{c}{\textbf{PA}$\uparrow$}
& \multirow{2}{*}{\textbf{VLM}$\uparrow$}
& \multicolumn{3}{c}{\textbf{NA}$\uparrow$}
& \multicolumn{3}{c}{\textbf{PA}$\uparrow$}
& \multirow{2}{*}{\textbf{VLM}$\uparrow$} \\
\cmidrule(lr){2-4} \cmidrule(lr){5-7}
\cmidrule(lr){9-11} \cmidrule(lr){12-14}
& \emph{Prec.} & \emph{Recal.} & \emph{F1}
& \emph{Prec.} & \emph{Recal.} & \emph{F1} &
& \emph{Prec.} & \emph{Recal.} & \emph{F1}
& \emph{Prec.} & \emph{Recal.} & \emph{F1} & \\

\midrule
\textit{Image-based Methods}\\
Nano~\cite{google2025gemini3proimage}
   &35.1&40.0&36.9&22.3&14.3&16.1&55.7 
  &38.3&39.2&38.1&22.8&17.9&18.2&59.2\\
PaperBanana~\cite{zhu2026paperbanana}
    &46.7&47.1&46.3&37.7&22.8&26.9&60.3
    &52.6&44.0&47.2&39.1&22.2&28.7&66.6\\
AutoFigure~\cite{zhu2026autofigure}
  &32.8&37.4&34.5&34.3&20.9&23.9&53.1
  &42.4&40.3&39.9&38.6&23.5&27.1&64.2\\
\midrule
\textit{SVG-based Methods}\\
GPT-5~\cite{openai2026gpt5}
  &35.6&21.6&25.6&27.3&10.1&13.8&31.1 
  &39.4&36.3&37.5&23.5&11.2&14.4&36.2 \\
TikZero~\cite{belouadi2025tikzero} 
   &20.4&11.2&14.9&9.6&1.7&2.3  & 4.2 
   &23.3&9.4&13.2&13.2&3.0&3.2&8.4\\
StarVector~\cite{rodriguez2025starvector} 
  &14.7&10.5&11.4&8.1&2.0&2.7&4.5
  &16.6&8.8&10.7&11.8&2.4&3.1&7.2 \\
Paper2SysArch~\cite{guo2025paper2sysarch} 
  &42.6&30.9&34.8&26.1&18.8&20.3&35.2 
 &39.1&37.5&36.3&22.5&19.7&19.4&39.3\\

Nano+EditBanana~\cite{google2025gemini3proimage}
  &34.6&36.1&35.3&18.4&11.9&12.6&50.4
  &37.2&37.0&35.7&17.6&13.5&14.3&56.8  \\
PaperBanana+EditBanana~\cite{zhu2026paperbanana}
  &44.9&43.6&44.5&31.0&18.4&21.2&55.9
  &50.9&39.6&43.8&31.6&17.5&20.8&58.7 \\
AutoFigure+EditBanana~\cite{zhu2026autofigure}
  &32.0&34.4&33.2&27.0&18.7&21.3&51.6
  &41.0&37.4&38.0&34.3&20.2&24.0&54.3\\
\hline
FigAgent (Ours)
  &\textbf{53.7}&\textbf{51.9}&\textbf{51.4}
  &\textbf{40.9}&\textbf{28.5}&\textbf{32.6}&\textbf{61.4}
  &\textbf{54.8}&\textbf{50.5}&\textbf{51.6}
  &\textbf{44.1}&\textbf{33.7}&\textbf{36.7}&\textbf{68.9} \\

\bottomrule
\end{tabular}
}
\label{tab:general}
\end{wraptable}
\noindent \textbf{Results.} 
\cref{tab:general} shows the comparative results. 
Among image-based methods, by using LLMs to parse paper texts, PaperBanana and AutoFigure achieve better performance compared to the simpler Nano baseline. However, these methods still suffer from low PA scores due to the inherent fidelity limitations of the underlying image generation models, which frequently produce visually appealing but structurally incorrect outputs.
On the other hand, we find that SVG-based methods such as StarVector and TikZero, despite producing editable SVG outputs, still underperform image-based methods across most metrics. 
This is mainly because these methods are specially designed for icons and simple data charts, and still struggle to adapt to more complex MI generation, which requires accurately depicting intricate visual components and their compositional relationships. We also find that Paper2SysArch, although achieving relatively good results on layout-related metrics (NA and PA), still exhibits notable deficiencies in the visual quality (as reflected by VLM Score).

Differently, by constructing drawing middlewares to support complex concept rendering and by guiding MI generation through the Explore-and-Select strategy that evaluates long-term impacts of drawing choices, our framework achieves the best performance across all metrics.
Qualitative comparisons are shown in~\cref{fig:vis}.

\subsection{Framework Analysis}
\label{sec:abalation}

Below, we conduct additional experiments on the FigAgentBench to further analyze the key characteristics of our framework. More results, including human evaluation, are in~\supp{Supplementary}.

\setlength{\columnsep}{0.05in}
\begin{wraptable}[6]{r}{0.5\columnwidth}
\centering
\vspace{-.8cm}
\caption{Evaluation on main mechanisms.}
\resizebox{\linewidth}{!}{
\begin{tabular}{l ccc ccc c}
\toprule
\multirow{2}{*}{\textbf{Method}} 
& \multicolumn{3}{c}{\textbf{NA}$\uparrow$} 
& \multicolumn{3}{c}{\textbf{PA}$\uparrow$} 
& \multirow{2}{*}{\textbf{VLM}$\uparrow$} \\
\cmidrule(lr){2-4} \cmidrule(lr){5-7}
& \emph{Prec.} & \emph{Recal.} &\emph{F1} & \emph{Prec.} & \emph{Recal.} &\emph{F1} & \\
\midrule
w/o middleware           & 29.8 & 24.3 & 26.5 & 24.7 &18.3&19.9 &39.5 \\
w/o Explore-and-Select       & 39.3 & 31.5 & 33.6 &32.4 & 20.2 & 23.8 &48.2 \\
w/ Greedy-Selection       & 43.7 & 43.2 & 41.9 &34.9 & 27.0 & 29.1 &54.3 \\
Ours&\textbf{53.7}&\textbf{51.9}&\textbf{51.4}
  &\textbf{40.9}&\textbf{28.5}&\textbf{32.6}&\textbf{61.4} \\
\bottomrule
\end{tabular}
}
\label{tab:wo main}
\end{wraptable}

\noindent\textbf{Impact of Main Mechanisms of FigAgent.}
We test the following variants:
\textbf{1) w/o middleware}, where the DA directly generates XML code segment for rendering each concept without using middleware;
\textbf{2) w/o Explore-and-Select}, where we remove our Explore-and-Select strategy, i.e., the DA renders the concepts in $\mathcal{G}$ one by one in a single pass, without trial-and-error drawing;
\textbf{3) w/ Greedy-Selection}, where we replace the Explore-and-Select strategy with a greedy mechanism: at each step, the DA selects the drawing choice with the highest score assessed by the EA.
As shown in~\cref{tab:wo main}, both the middleware repository and the Explore-and-Select strategy contribute significantly to overall efficacy.

\setlength{\columnsep}{0.05in}
\begin{wraptable}[8]{r}{0.5\columnwidth}
\centering
\vspace{-.8cm}
\caption{Evaluation on middleware evolution mechanism.}
\resizebox{\linewidth}{!}{
\begin{tabular}{l ccc ccc c}
\toprule
\multirow{2}{*}{\textbf{Method}} 
& \multicolumn{3}{c}{\textbf{NA}$\uparrow$} 
& \multicolumn{3}{c}{\textbf{PA}$\uparrow$} 
& \multirow{2}{*}{\textbf{VLM}$\uparrow$} \\
\cmidrule(lr){2-4} \cmidrule(lr){5-7}
& \emph{Prec.} & \emph{Recal.} &\emph{F1} & \emph{Prec.} & \emph{Recal.} &\emph{F1} & \\
\midrule

w/o evolution  & 37.3 & 27.2 & 31.8 & 30.2 & 19.7 & 22.9 &45.5 \\
w/o selection    &40.1 &30.7&33.9  &32.1& 20.3 &24.2 &47.6\\
w/o mutation  & 48.9 & 46.6 & 46.8 &36.2 & 27.1 &29.5 &59.3\\
w/o crossover  &46.5&44.3 &43.8    &35.8 & 25.8& 28.3 &54.2\\
Ours &\textbf{53.7}&\textbf{51.9}&\textbf{51.4}
  &\textbf{40.9}&\textbf{28.5}&\textbf{32.6}&\textbf{61.4} \\
\bottomrule
\end{tabular}
}
\label{tab:evolution}
\end{wraptable}
\noindent\textbf{Impact of Middleware Evolution Mechanism.}
We compare the following variants:
\textbf{1) w/o evolution}, where the middleware evolution process is entirely removed, and the DA can only invoke middlewares from the initially created repository for drawing MIs;
\textbf{2) w/o selection}, \textbf{3) w/o mutation}, and \textbf{4) w/o crossover}, where the corresponding operation in the evolution process is removed.
As shown in~\cref{tab:evolution}, our full framework outperforms all ablated variants.
Specifically, selection \textit{removes} low-quality middlewares, mutation \textit{refines} existing middlewares, and crossover \textit{explores new} middleware designs, all of which contribute positively to the efficacy of middleware evolution.

\noindent\textbf{Scaling to Evolving Research Community.}
To validate our framework's scalability, we first build $\mathcal{M}$ using MI resources in $\mathcal{D}_{\text{exp}}$ published only up to Dec 2023, then examine following variants, all evaluated on the full test set:
\textbf{Ours (2023)}: using the 2023-based $\mathcal{M}$ directly for evaluation;
\textbf{Ours (2023 direct adding)}: new middlewares created from Dec 2023--Dec 2025 MI resources are directly added to the 2023-based $\mathcal{M}$ without middleware evolution;
\textbf{Ours (2023 evolution)}: new middlewares from Dec 2023--Dec 2025 MI resources are integrated into $\mathcal{M}$ through middleware evolution; and
\textbf{Ours (2025)}: $\mathcal{M}$ is constructed from scratch using all MI resources published up to Dec 2025.
\setlength{\columnsep}{0.05in}
\begin{wraptable}[7]{r}{0.5\columnwidth}
\centering  
\vspace{-.6cm}
\caption{Evaluation on scalability to the evolving research community.}
\resizebox{\linewidth}{!}{
\begin{tabular}{l ccc ccc c}
\toprule
\multirow{2}{*}{\textbf{Method}} 
& \multicolumn{3}{c}{\textbf{NA}$\uparrow$} 
& \multicolumn{3}{c}{\textbf{PA}$\uparrow$} 
& \multirow{2}{*}{\textbf{VLM}$\uparrow$} \\
\cmidrule(lr){2-4} \cmidrule(lr){5-7}
& \emph{Prec.} & \emph{Recal.} &\emph{F1} & \emph{Prec.} & \emph{Recal.} &\emph{F1} & \\
\midrule

Ours (2023) &39.8&36.4&36.3&32.2&25.3&26.4&49.5 \\
Ours (2023 direct adding)  &42.0&36.5&37.4&35.9&26.2&28.1&56.8 \\
Ours (2023 evolution) &54.3&51.8&50.9&40.5&27.7&31.2&64.9 \\
Ours (2025) &53.7&51.9&51.4
  &40.9&28.5&32.6&61.4 \\

\bottomrule
\end{tabular}
}
\label{tab:scale}
\end{wraptable}
As shown in~\cref{tab:scale},
Ours (2023),  with relatively outdated $\mathcal{M}$, struggles to handle new concepts emerging in the research community, and Ours (2023 direct adding) still performs limited, mainly because adding middlewares without evolution introduces low-quality and redundant middlewares, degrading the overall efficacy of $\mathcal{M}$.
In contrast, Ours (2023 evolution) achieves performance close to Ours (2025), while the latter requires rebuilding the repository from scratch, making it less suitable for the rapidly evolving research community,
but our evolution mechanism operates only at the middleware level without touching any model parameters, enabling efficient updates. More analysis of efficiency and performance at different update frequencies are in~\supp{Supplementary}.

\vspace{-0.2cm}

\section{Conclusion}

We propose FigAgent, a novel multi-agent framework for MI generation. By distilling drawing experiences from published scientific papers to create and evolve a drawing middleware repository, and by performing drawing under a novel Explore-and-Select strategy that evaluates the long-term impact of each drawing choice, our framework automatically generates high-fidelity SVG MIs for scientific papers. Extensive experiments demonstrate the efficacy of our method.

\bibliography{main}
\bibliographystyle{plain}

\end{document}